\documentclass[prl,twocolumn,longbibliography,amsmath,superscriptaddress,amssymb,nobibnotes]{revtex4-2}

\usepackage{graphicx}
\usepackage{dcolumn}
\usepackage{bm}
\usepackage{amssymb}
\usepackage{amsmath}
\usepackage{color}
\usepackage[dvipsnames]{xcolor}
\usepackage{multirow}
\usepackage{physics}
\usepackage{mhchem}
\usepackage{chemformula}
\usepackage{booktabs}
\usepackage{amsmath} 
\usepackage[T2A]{fontenc}
\usepackage{chemformula}
\usepackage{siunitx} 

\usepackage[linkcolor=blue,urlcolor=blue,citecolor=blue,colorlinks=true]{hyperref}

\usepackage{xr}
\usepackage{siunitx}
\usepackage[normalem]{ulem}

\begin{document}
\title{Cation-intermixing driven evolution from a narrow avoided crossing to \\ an enlarged inverted gap in Ge$_2$Bi$_2$Te$_5$}

\author{Alexander~S.~Frolov}
\email{frolov.as@mipt.ru}

\affiliation{Moscow Center for Advanced Studies, Kulakova str. 20, Moscow 140829, Russia}
\affiliation{Donostia International Physics Center (DIPC), 20018 Donostia/San Sebasti\'{a}n, Basque Country, Spain}
\affiliation{Lomonosov Moscow State University, Leninskie Gory 1/3, 119991 Moscow, Russia }

\author{Alexander~V.~Fedorov}
\affiliation{Helmholtz-Zentrum Berlin f\"ur Materialien und Energie, Elektronenspeicherring BESSY II, Albert-Einstein-Str. 15, 12489 Berlin, Germany}

\author{Nadezhda~V.~Vladimirova}
\affiliation{Moscow Center for Advanced Studies, Kulakova str. 20, Moscow 140829, Russia}
\affiliation{Lomonosov Moscow State University, Leninskie Gory 1/3, 119991 Moscow, Russia }

\author{Sergey~M.~Aksenov}
\affiliation{Laboratory of Arctic Mineralogy and Materials Sciences, FRC Kola Science Center, 14 Fersman Str. Apatity, Murmansk Region, 184209 Russian Federation}

\author{Mark~A.~Naumov}
\affiliation{Moscow Center for Advanced Studies, Kulakova str. 20, Moscow 140829, Russia}

\author{Savely~V.~Domenko}
\affiliation{Moscow Center for Advanced Studies, Kulakova str. 20, Moscow 140829, Russia}

\author{Artem~A.~Minaev}
\affiliation{Moscow Center for Advanced Studies, Kulakova str. 20, Moscow 140829, Russia}

\author{Vasily~S.~Stolyarov}
\affiliation{Moscow Center for Advanced Studies, Kulakova str. 20, Moscow 140829, Russia}

\author{Denis~V.~Vyalikh}
\email{denis.vyalikh@dipc.org}

\affiliation{Donostia International Physics Center (DIPC), 20018 Donostia/San Sebasti\'{a}n, Basque Country, Spain}
\affiliation{University of the Basque Country UPV/EHU, 20018 Donostia/San Sebastián, Spain}
\affiliation{IKERBASQUE, Basque Foundation for Science, 48013, Bilbao, Spain}

\author{Lada~V.~Yashina}
\affiliation{Moscow Center for Advanced Studies, Kulakova str. 20, Moscow 140829, Russia}
\affiliation{Donostia International Physics Center (DIPC), 20018 Donostia/San Sebasti\'{a}n, Basque Country, Spain}
\affiliation{Lomonosov Moscow State University, Leninskie Gory 1/3, 119991 Moscow, Russia }

\date{\today}

\begin{abstract}

Combining the building blocks and structural motifs of the topological insulator \ch{Bi2Te3} and the ferroelectric Rashba semiconductor \ch{GeTe}, we synthesized and comprehensively characterized the resulting \ch{Ge2Bi2Te5} compound, which exhibits pronounced cation intermixing. Using momentum-resolved photoemission spectroscopy and \textit{ab initio} calculations, we demonstrate how cation intermixing drives the evolution of its electronic structure from a narrow avoided crossing to a substantially enlarged inverted band gap.  We establish the microscopic origin of this behavior in the redistribution of Ge and Bi atoms within the cation sublattice, which increases the contribution of Bi-derived states near the valence-band maximum and enhances the effective contribution of spin–orbit coupling, thereby stabilizing the topological electronic structure.  Our findings demonstrate that cation disorder, which can be controlled during crystal growth, provides an additional degree of freedom for engineering topological electronic structures, opening new opportunities for the development of functional quantum materials.

\end{abstract}


\maketitle

\section{Introduction}
Topological insulator materials (TIs), discovered about two decades ago, have continuously attracted considerable interest in the research community due to their rich fundamental physics and promising potential for applications across various technological fields~\cite{Hasan2010, Qi2011, Tokura2019}. Topological insulators are characterized by a bulk energy gap induced by spin–orbit coupling (SOC)~\cite{Zhang2009}. At the surface of such crystals, this leads to the emergence of gapless states with Dirac-like dispersion, the so-called topological surface states (TSSs), which reveal a peculiar spin texture and give rise to robust conducting channels~\cite{Chen2009,Hsieh2009}. While these materials have been extensively investigated over the years, essential questions remain unresolved, in particular regarding how the topological phase emerges and how to develop reliable strategies for identifying new materials with such remarkable properties. One approach to address this problem is to trace the topological phase transition $via$ compositional variation in systems combining topological and trivial compounds, thereby tuning the strength of SOC~\cite{Xu2011, Guo2011}.

One representative system in which this approach can be realized is the GeTe -- Bi$_2$Te$_3$ system, which combines structural blocks of the prototypical topological insulator Bi$_2$Te$_3$ and the ferroelectric Rashba semiconductor GeTe exhibiting giant Rashba SOC~\cite{matsunaga2007structures-9e4, DiSante2013,Liebmann2016}.



In the GeTe–Bi$_2$Te$_3$ system, compositional variation leads to the formation of several distinct ternary compounds with well-defined layered structures~\cite{matsunaga2007structures-9e4,karpinsky1998x-ray-57a,Alakbarova2022}. Note that in solid solutions, many physical properties can be finely tuned by adjusting the composition. However, the associated chemical disorder increases configurational entropy due to statistical intermixing and can lead to enhanced carrier scattering ~\cite{konstantinov1999transport-767}. In contrast, the individual multicomponent phases in the GeTe–Bi$_2$Te$_3$ system possess distinct layered crystal structures, providing an opportunity to systematically investigate the evolution of their electronic and topological properties with composition. Of particular interest is the vicinity of the \ch{Ge2Bi2Te5} phase, where changes in the electronic structure may drive the system toward a topological phase transition.

In this work, we demonstrate that the degree of cation intermixing provides an effective control parameter for tuning the electronic structure of Ge$_2$Bi$_2$Te$_5$. We focus on the structural properties of the GeTe–Bi$_2$Te$_3$ family systems and their electronic structure. Single crystals were synthesized both from the melt and by chemical vapor transport, followed by comprehensive structural characterization using X-ray diffraction (XRD), high-angle annular dark-field scanning transmission electron microscopy combined with energy-dispersive X-ray spectroscopy (HAADF-STEM/EDX), and atomic force microscopy (AFM). The electronic band structure was investigated by momentum-resolved photoemission spectroscopy (ARPES), supported by complementary \textit{ab initio} calculations.
Our results show that crystals grown from the melt and by chemical vapor transport exhibit different degrees of Ge/Bi cation intermixing. This difference is quantified by single-crystal X-ray diffraction and independently supported by atomically resolved microscopy measurements. A combined analysis of these structural results, ARPES data, and first-principles calculations demonstrates that increasing cation intermixing drives the evolution of the band structure from a narrow avoided crossing along $\Gamma$--$Z$ to an enlarged inverted gap characteristic of a topological insulator.

\section{Results and Discussion}

\subsection{1. Atomic structure of \ch{Ge2Bi2Te5}}

We begin with the structural analysis of the synthesized crystals, the results of which are summarized in Figure~\ref{fig_1}, focusing on two types of samples grown from vapor and from melt. As shown below, these samples exhibit different degrees of cation intermixing and thus provide a controllable parameter for tuning the electronic properties. Figures~\ref{fig_1}a,b show photographs of the melt- and vapor-grown crystals, respectively, with the millimeter graduations on the scale visible at the bottom. Large single crystal with homogeneous composition (see Supporting Information, Fig.~S1) and high structural quality was obtained from the melt, whereas vapor growth resulted in smaller crystals grown without contact with the ampoule walls, i.e. free from mechanical stress and the corresponding dimensional imperfections.

The HAADF-STEM image of the crystal cross-section for the vapor-grown sample, shown in Figure~\ref{fig_1}d, reveals a highly ordered structure composed of nonuple layers (NLs). The NL structure is further confirmed by AFM measurements of the freshly cleaved Ge$_2$Bi$_2$Te$_5$ surface (Fig.~\ref{fig_1}f). The surface exhibits extended terraces separated by steps of uniform height $h=1.7$ nm, consistent with the NL thickness and in good agreement with the XRD result ($h=c$). Next, Fig.~\ref{fig_1}g shows the STM image of the vacuum-cleaved surface obtained with atomic resolution. Despite the well-ordered hexagonal atomic structure, the electronic landscape exhibits pronounced spatial variations caused by point defects located in the underlying layers, most likely associated with cation antisite defects.

To unveil the cation intermixing, we focus on the high resolution EDX maps shown in Fig.~\ref{fig_1}e, which reveal atomically resolved patterns integrated over a selected area. Together with these, the corresponding HAADF-STEM image (left) is shown as a reference. Green corresponds to Te atoms occupying the anion sublattice. In contrast to the ideal structure, where Bi resides in the outer cation layers and Ge in the inner ones, all cation sites are occupied by both Ge and Bi, indicating pronounced intermixing. This is particularly evident in the Bi map (blue), where four atomic rows are clearly resolved. At the same time, the outer layer exhibits higher intensity, pointing to a preferential occupation of the outer cation layer by the heavier Bi atoms.

Quantitative evaluation of the site occupancies, derived from single-crystal XRD, is summarized in Table~\ref{tab:Cryst_1}. The degree of disorder is characterized by the ordering factor $\alpha$, which equals unity for an ideally ordered crystal and approaches zero for completely intermixed cation sites. The degree of cation ordering is quantified by
\begin{equation}
    \alpha = 
\frac{x_{\mathrm{Ge}}^{\mathrm{M1}}
           - x_{\mathrm{Ge}}^{\mathrm{M2}}}{x_{\mathrm{Ge}}^{\mathrm{M1}}
           + x_{\mathrm{Ge}}^{\mathrm{M2}}}
           ,
\end{equation}
where $x_{\mathrm{Ge}}^{\mathrm{M1}}$ and $x_{\mathrm{Ge}}^{\mathrm{M2}}$ are the Ge occupancies of the inner and outer cation sites, respectively. The limit $\alpha=1$ corresponds to ideal cation ordering, whereas $\alpha\rightarrow0$ corresponds to complete intermixing. 

As seen for both samples, a high level of disorder is observed, with a slightly larger degree of disorder in the vapor-grown crystal. For this sample, the Ge/Bi atomic ratio deviates from unity, reaching 1.045, suggesting that this imbalance may promote additional disorder. In principle, this effect could be controlled by optimizing the preparation protocol and thus may serve as a tunable parameter governing the resulting electronic and topological properties.

We note that the Bi/Ge intermixing is repeatedly observed for such layered compounds~\cite{karpinsky1998x-ray-57a, matsunaga2007structures-9e4, zhang2021cation-dfb}. This behavior is not surprising, given the similarity of the covalent radii of Bi (146~pm) and Ge (122~pm), as well as their nearly identical electronegativities (2.01 and 2.02, respectively).

\begin{figure*}[t]
\includegraphics[width=0.99\linewidth]{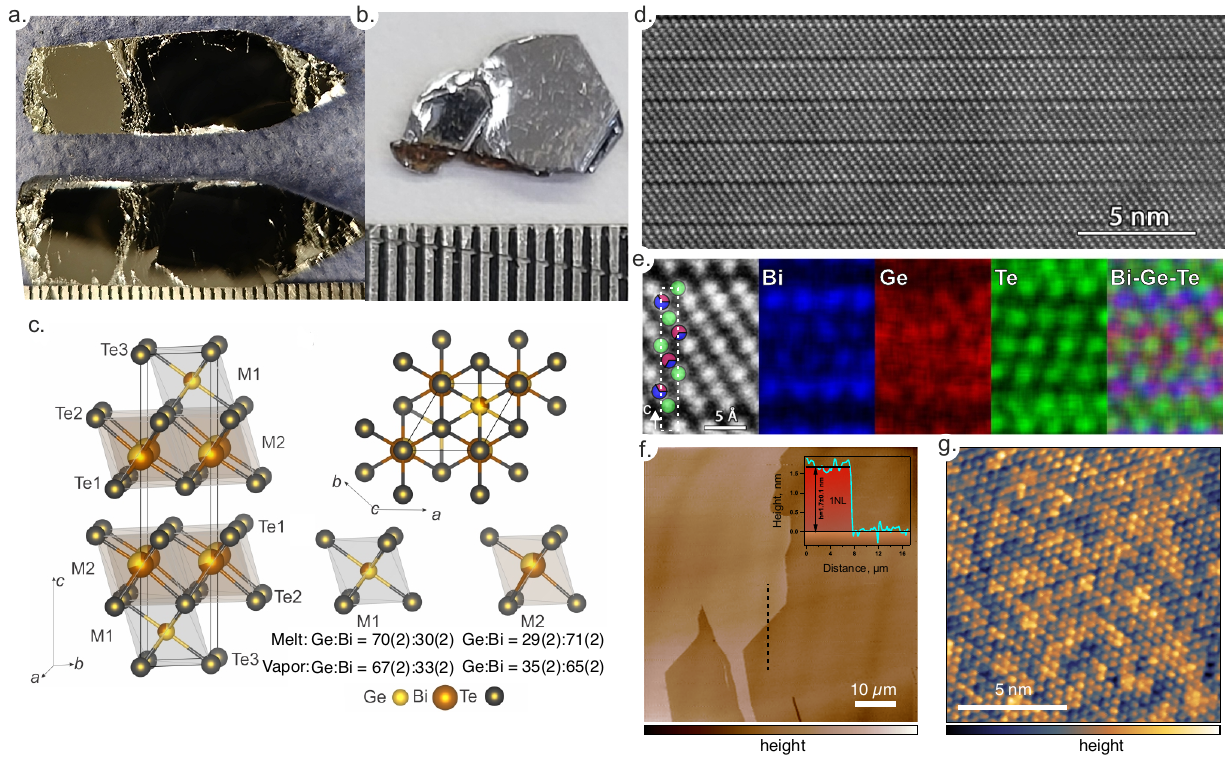}
\caption{\textbf{Structural characterization of Ge$_2$Bi$_2$Te$_5$ crystals.} Photographs of samples grown (a) from the melt and (b) by chemical vapor transport. (c) Unit cell, projection onto the $ab$ plane, and coordination polyhedra of the inner M1 and outer M2 cation sites. (d) HAADF-STEM image viewed along the [100] zone axis. (e) Averaged STEM/EDX elemental maps. (f) AFM image of a vacuum-cleaved surface; the inset shows the corresponding height profile. (g) STM image of a vacuum-cleaved surface acquired at 4~K.}
\label{fig_1}
\end{figure*}

\begin{table*}[]

\caption{Chemical composition and cation site occupancies of Ge$_2$Bi$_2$Te$_5$ single crystals grown by the modified Bridgman method \cite{sergeev2025growth-648} and CVT. The chemical compositions were determined by SC-XRD and XRF, while the cation site occupancies were obtained from SC-XRD measurements at 296 K. The ordering parameter $\alpha$ is defined in Eq. (1). Values in parentheses indicate one standard uncertainty in the last digit(s).}

\label{tab:Cryst_1}
\renewcommand{\arraystretch}{1.1}
\begin{tabular}{|ccc|c|c|}
\hline
\multicolumn{3}{|c|}{Growth method}                                     & Modified Bridgman method    & CVT                         \\ \hline
\multicolumn{2}{|c|}{\multirow{2}{*}{Composition}}                                                               & SC XRD & Ge$_{1.99(6)}$Bi$_{2.01(6)}$Te$_5$ & Ge$_{2.04(7)}$Bi$_{1.96(7)}$Te$_5$ \\ \cline{3-5} 
\multicolumn{2}{|c|}{}                                            & XRF & Ge$_{1.99(1)}$Bi$_{2.01(1)}$Te$_{5.00(2)}$ & Ge$_{2.05(1)}$Bi$_{1.96(1)}$Te$_{4.99(2)}$ \\ \hline
\multicolumn{1}{|c|}{\multirow{4}{*}{\begin{tabular}[c]{@{}c@{}}Occupation \\ for cation \\ position\end{tabular}}} &
  \multicolumn{1}{c|}{\multirow{2}{*}{M1}} & Ge1    & 0.70(2)               & 0.67(2)               \\ \cline{3-5} 
\multicolumn{1}{|c|}{} & \multicolumn{1}{c|}{}                    & Bi1 & 0.30(2)                     & 0.33(2)                     \\ \cline{2-5} 
\multicolumn{1}{|c|}{} & \multicolumn{1}{c|}{\multirow{2}{*}{M2}} & Ge2 & 0.29(2)                     & 0.35(2)                     \\ \cline{3-5} 
\multicolumn{1}{|c|}{} & \multicolumn{1}{c|}{}                    & Bi2 & 0.71(2)                     & 0.65(2)                     \\ \hline
\multicolumn{2}{|c|}{degree of ordering}                          & $\alpha$   & 0.41(1)                     & 0.31(1)                     \\ \hline
\end{tabular}
\vspace{-10pt}
\end{table*}

\subsection{2. Electronic band structure probed by ARPES}

\begin{figure*}[t]
\includegraphics[width=0.99\linewidth]{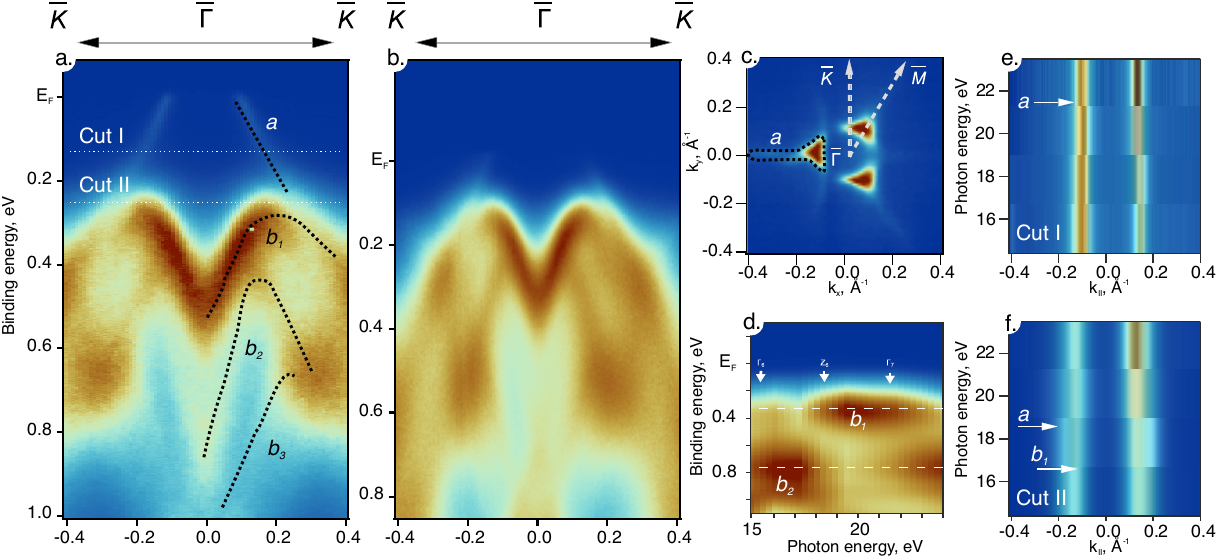}
\caption{\textbf{ARPES insights into the electronic structure of Ge$_2$Bi$_2$Te$_5$}. ARPES spectra of the melt-grown (a) and vapor-grown (b) Ge$_2$Bi$_2$Te$_5$ crystals acquired at a photon energy of 19~eV. Fermi-surface map measured at 19~eV for vapor-grown crystal  (c). Photon-energy dependence of the ARPES intensity taken along the $\overline{K}$--$\overline{\Gamma}$--$\overline{K}$ direction (d). (e,f) Photon-energy-dependent ARPES intensity taken along cuts~I and~II indicated in (a), respectively. }
\label{Fig_2}
\end{figure*}

To explore the electronic structure of Ge$_2$Bi$_2$Te$_5$, we performed ARPES measurements on the melt-grown and vapor-grown crystals introduced above. The corresponding electronic band structures are shown in Fig.~\ref{Fig_2}a,b. While both samples are hole-doped, the vapor-grown crystal exhibits a somewhat lower doping level.  In each case, the Fermi level ($E_F$) intersects the valence band close to its maximum. In GeTe-based materials, hole doping is commonly attributed to metal vacancies~\cite{frolov2020atomic-15b}. For mixed GeTe–Bi$_2$Te$_3$ compounds, positively charged \ch{Ge_{Bi}} antisite defects have also been proposed as an important source of hole carriers~\cite{konstantinov1999transport-767,shelimova2004crystal-b12}. Our structural and chemical characterization reveals substantial cation intermixing together with a slight excess of Ge (Table~\ref{tab:Cryst_1}), which points to the presence of a substantial concentration of uncompensated Ge$_{\mathrm{Bi}}^{\prime}$ antisite defects and the observed $p$-type doping.

\begin{table}[h!]
\centering
\caption{Transport properties of crystals}
\label{tab:t}
\begin{tabular}{@{}lcc@{}}
\toprule
Parameters & Bridgman & CVT \\
\midrule
\(p\), \(10^{19} \, \mathrm{cm}^{-3}\)
    & \(6.7\)
    & \(28.6\) \\
\(\mu\), \(\mathrm{cm}^{2}/(\mathrm{V}\cdot\mathrm{s})\)
    & \(108\)
    & \(73\) \\
\(\Theta_{\mathrm{D}}^{*}\), \(\mathrm{K}\)
    & \(131.74\)
    & \(135.74\) \\
\bottomrule
\end{tabular}

\smallskip

\end{table}

For both samples, the upper part of the valence band is formed by the $a$ bands located near the Fermi level at $k < \SI{0.4}{\per\angstrom}$ along the $\overline{\Gamma}$--$\overline{K}$ direction of the surface Brillouin zone in (Fig.~\ref{Fig_2}a,b). Although these states look similar to the topological surface state, they exhibit a pronounced threefold symmetry, as clearly seen in the Fermi surface map shown in Fig.~\ref{Fig_2}c, indicating their bulk origin. Upon moving to larger momenta, the $a$ band crosses the $b_1$ band at $k \approx \SI{0.28}{\per\angstrom}$.

In the high binding-energy (BE) region of the valence band, a series of intense hole-like features, denoted as $b_m$, is observed (see Fig.~\ref{Fig_2}). Note that in tetradymite-like crystals, such states are predominantly derived from Te $5p_z$ orbitals, as is common for this family of materials~\cite{Zhang2009}.

To understand the formation of the $b_m$ states, we consider the following picture. In the absence of chemical bonding and crystal-field effects, the $b_m$ states can be viewed as originating from atomic-like Te $p$ states. Then, the total degeneracy is given by $2 \cdot 3 \cdot l$, where $l$ denotes the number of non-equivalent tellurium positions, and the factors 2 and 3 arise from the spin and orbital degrees of freedom, respectively ($l$=3 for nonuple layers). 

Upon the formation of chemical bonds and inclusion of the crystal field, this degeneracy is lifted. First, the symmetry of the crystal environment removes the equivalence between different Te sites. Second, the orbital degeneracy is lifted, leading to a splitting into $p_{x,y}$ and $p_z$ components. As a result, a manifold of valence-band states emerges, with the number of branches primarily determined by the number of crystallographically inequivalent Te layers. In Ge$_2$Bi$_2$Te$_5$, the nonuple-layer structure contains three such Te layers, giving rise to the observed multiplicity of the $b_m$ states.

The Te $p_{x,y}$ derived states are located at higher BEs, more than 1 eV below ($E_F$). Within this approach, each ($b_m$) state can be associated with a particular Te layer and is predominantly derived from Te ($p_z$) orbitals, reflecting the underlying crystallographic symmetry. Although this description is qualitative and neglects the details of chemical bonding, it captures the essential physics. In particular, the discussed $b_m$ states are expected to exhibit a pronounced quasi-two-dimensional character, consistent with the observed splitting between them, while the index $m$ labels the crystallographically non-equivalent Te sites.

Two-dimensional character of $b_m$ states is indeed revealed in the absence of energy dispersion in $k_z$ direction, which we probe by photon energy variation.  Figure~\ref{Fig_2}d shows photon energy dependence of bands taken at the $\overline{\Gamma}$ point. Distinct intensity increases are observed at binding energies of approximately 0.45~eV and 0.85~eV. These features exhibit a periodic decaying pattern as a function of photon energy and are associated with the $b_m$ states discussed earlier. The periodic dependence of the state intensities, along with the absence of notable energy dispersion with photon energy, indicates that these states exhibit surface resonance behavior. Even more clearly 2D character is revealed in Figures~\ref{Fig_2}e,f where photon energy dependence is shown for the cut~I and cut~II of Figure~\ref{Fig_2}a, which are relevant to the bands $a$ and $b_m$ correspondingly. The observed features exhibit a pronounced quasi-two-dimensional character, manifested by the complete absence of dispersion along the $k_z$ direction.

In our ARPES experiment, the electron inelastic mean free path, estimated using the TPP-2M formalism, is approximately ($\lambda \approx \SI{7}{\angstrom}$). As a result, nearly all detected photoelectrons originate from a depth of (3$\lambda \approx \SI{21}{\angstrom}$), which is comparable to the thickness of a Ge$_2$Bi$_2$Te$_5$ nonuple layer (\SI{17.343}{\angstrom}). Thus, the measurements predominantly probe the electronic structure of the surface nonuple layer, including possible contributions from surface-resonance states.

\subsection{3. Influence of cation intermixing on the electronic band structure}

\begin{figure*}[t]
\includegraphics[width=0.99\linewidth]{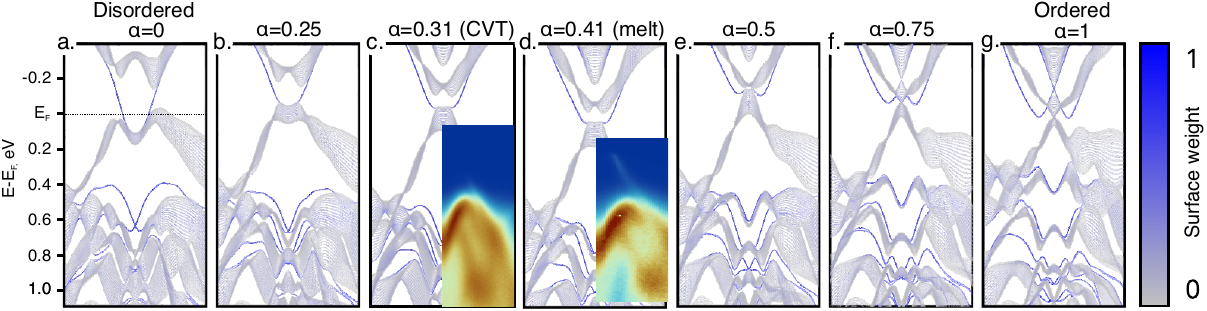}
\caption{\textbf{Theoretical insight into the band structure of Ge$_2$Bi$_2$Te$_5$}. Evolution of the electronic structure with increasing cation disorder. Calculated bulk-projected and surface spectral functions for different values of the ordering parameter $\alpha$. Blue denotes the surface spectral weight. Experimental ARPES patterns are overlaid in panels (d) and (e) to facilitate comparison with the calculations.}
\label{fig_3}
\end{figure*}

Since the studied crystals are $p$-doped, the experimental ARPES pattern alone does not allow us to determine whether the $\mathrm{Ge_2Bi_2Te_5}$ material is in a topological phase. To clarify this point, we performed density functional theory (DFT) calculations, the results of which are shown in Fig.~\ref{fig_3}. We first considered an ideally ordered crystal structure ($\alpha=1$); left panel of Fig.~\ref{fig_3}, for which the calculated band structure fails to reproduce the experimental ARPES data. We therefore introduced cation intermixing, parameterized by the order parameter ($\alpha$) (ref.~\cite{vladimirova2023occupancy-fe6}), and found that the calculated band structure depends strongly on the degree of disorder.

Cation intermixing leads to an increase in the bulk band gap and substantial modifications of the surface-state dispersion. In the ideally ordered phase, the surface state exhibits a W-shaped dispersion resembling a Rashba-split parabolic band, with its Kramers point overlapping the bulk conduction band.
With increasing intermixing, the surface-state dispersion becomes more complex and evolves towards a Dirac-like cone with a finite parabolic contribution. In particular, intermixing changes the separation and curvature of the gap-edge bands and modifies the surface spectral function. Good agreement with experiment is obtained for $\alpha = 0.41\,$ and $\alpha = 0.31\,$, consistent with the value derived from SC-XRD. To establish this correspondence, we have used the energy positions of the main spectral features: position of the $b_1$ state at $k_{\parallel}=0$ and at the band maxima; energy gap between the $a$ and $b_1$ states at $k_{\parallel}=0$. Further details are provided in Fig. S2 of the Supporting Information. Our data also indicate that the intermixing is slightly different in vapor-grown and melt-grown crystals, which leads to small differences in their band structures. Experimental data are overlaid with calculated ones in Fig.~\ref{fig_3}d,e for comparison.

A comparison between the ARPES data and the calculated band structures allows us to conclude that both studied crystals are in the topological insulating regime. 
At the same time, the evolution of the surface-state and bulk dispersion does not imply a change in the global topological class. Wannier charge-center calculations yield $(\nu_0;\nu_1\nu_2\nu_3)=(1;000)$ for every considered value of $\alpha$. Thus, cation intermixing continuously reshapes the bulk and surface electronic structure within a single strong topological phase. This constitutes our central finding and establishes controlled cation disorder as an effective approach to engineering the electronic band structure in this class of materials.

\subsection{4. Site-resolved orbital character of the gap-edge states}

To gain further insight into the origin of this behavior, we calculated the partial density of states (PDOS) for selected crystallographic sites, allowing us to track the band inversion near the $\overline{\Gamma}$ point. The calculations were performed using the ordering parameter $\alpha = 0.41$, determined experimentally for the Bridgman-grown crystal. The chosen $k$-path covers all four symmetry-inequivalent time-reversal invariant momentum (TRIM) points, namely $\Gamma = (0,0,0)$, $Z = (0,0,0.5)$, $F = (0.5,0,0)$, and $L = (0.5,0.5,0)$.

\begin{figure*}[t]
\includegraphics[width=0.99\linewidth]{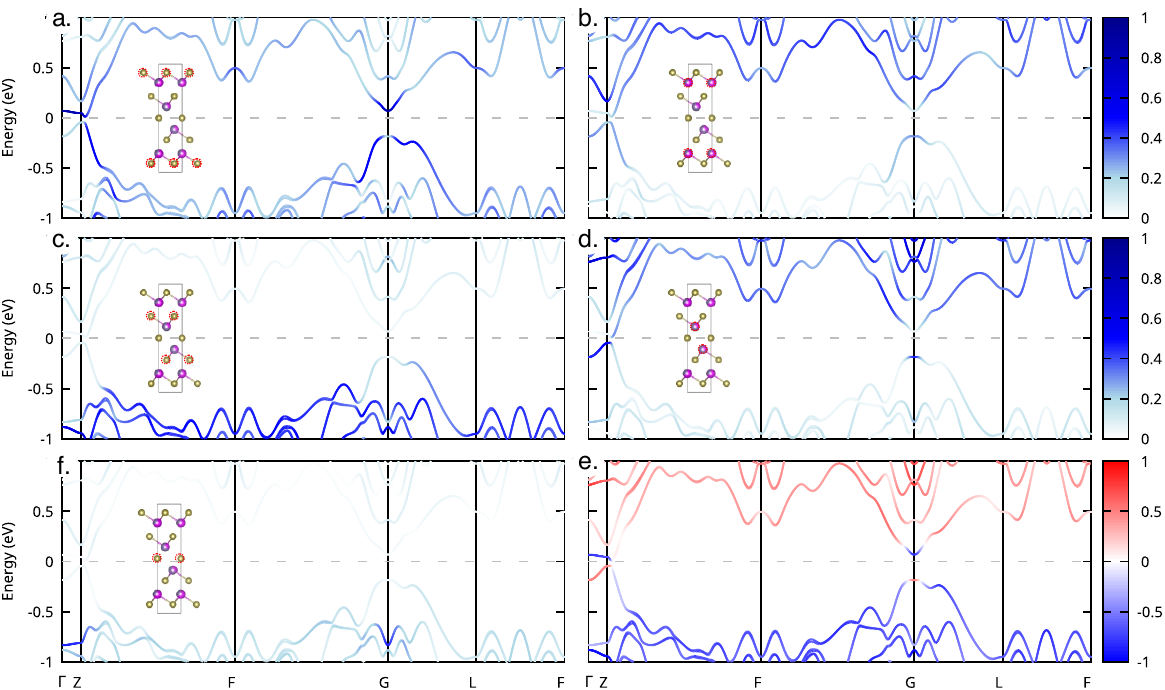}
\caption{\textbf{Site-resolved electronic structure of Ge$_2$Bi$_2$Te$_5$ at $\alpha$=0.41}. The color denotes spectral weight of corresponding atomic site (a) Te near VdW gap (Te$_1$, (b) Bi position (M2), (c) Te$_2$ position, (d) Ge position (M1), and (e) Central Te position (Te$_3$). (f) Difference between the total cation and anion orbital weights, $\Sigma(\mathrm{Ge+Bi})-\Sigma(\mathrm{Te})$.}
\label{fig_4}
\end{figure*}

Figures~\ref{fig_4}a--e display the orbital weight projected onto the five symmetry-inequivalent atomic sites in the \ch{Ge2Bi2Te5} lattice.  One can clearly see that the states at the edges of the bulk band gap, especially in the vicinity of the $\Gamma$ and $Z$ points, are predominantly composed of orbitals from two sites: the tellurium atoms near the Van der Waals gap (Te1) and the metal atoms located near the center of the NL block (M1). As $\alpha$ decreases, the Bi contribution at M1 increases and the calculated separation of the relevant band manifolds grows. This finding correlates with the mechanism underlying the observed enhancement of the calculated inverted band gap (Fig.~\ref{fig_3}): substituting Ge with Bi on the M1 site effectively increases the spin-orbit coupling acting on these gap-edge states, because the atomic number of Bi ($Z = 83$) is significantly larger than that of Ge ($Z = 32$). As a result, the cation intermixing acts as a highly efficient way to boost the effective spin--orbit coupling and to increase the band gap in the topological phase.

Figure~\ref{fig_4}f presents a differential plot illustrating the difference between the total cation (Ge and Bi) weight and the total anion (Te) weight, $\Sigma(\mathrm{Ge+Bi}) - \Sigma(\mathrm{Te})$. 
The redistribution of cation and anion weights characterizes the evolution of the orbital composition, but does not by itself determine inversion parity. The topological classification is obtained independently from Wannier charge-center evolution, which yields $(\nu_0;\nu_1\nu_2\nu_3)=(1;000)$ for every examined $\alpha$. The orbital-character reversal is therefore discussed as an evolution within the same strong topological phase.

Another essential observation is that the states near the valence-band maximum, corresponding to the $b_{1-3}$ features observed in ARPES, exhibit clearly different spectral weight distributions among the non-equivalent Te sites (Figs.~\ref{fig_4}a,c,e). The fact that they are predominantly composed of Te $p_z$ orbitals suggests that each of these subbands can be associated with a specific Te layer within the NL block, in agreement with the qualitative picture discussed above. Indeed, the site-projected band structure reveals that the individual $b_m$ features exhibit clearly distinct weight distributions across the inequivalent Te positions, supporting the assignment of the index $m$ to a particular crystallographic tellurium layer.

\subsection{5. Crossover from a narrow avoided crossing to an enlarged inverted gap driven by cation intermixing}

\begin{figure*}[t]
\includegraphics[width=0.99\linewidth]{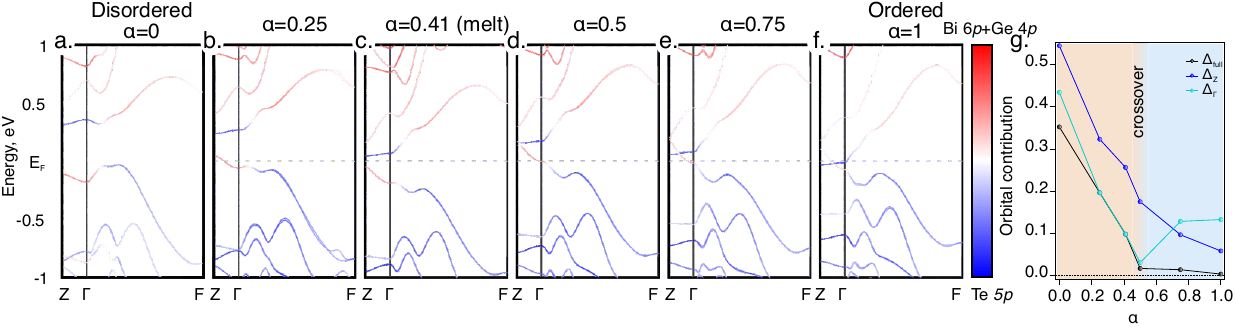}
\caption{\textbf{Cation-intermixing-induced evolution of the inverted band structure of Ge$_2$Bi$_2$Te$_5$.} (a--f) Site-projected bulk band structures calculated for different degrees of cation intermixing. (g) Evolution of the band-gap sizes at $\Gamma$ and $Z$ as a function of the ordering parameter $\alpha$.}
\label{fig_5}
\end{figure*}

Figure~\ref{fig_5} illustrates the evolution of the band structure as a function of the disorder parameter $\alpha$ along the high-symmetry path $Z$--$\Gamma$--$F$, with the line color encoding the orbital weight difference $\Sigma(\mathrm{Ge+Bi}) - \Sigma(\mathrm{Te})$. As the disorder increases from the fully ordered structure ($\alpha = 1$) towards the maximally intermixed case ($\alpha \rightarrow 0$), one can clearly observe a gradual growth of the inverted band gap at both the $\Gamma$ and $Z$ points. 
For $\alpha\gtrsim0.5$, the relevant valence- and conduction-band branches approach each other along $\Gamma$--$Z$ and form a narrow avoided crossing. The minimum local separation is only several meV, placing the ordered and weakly intermixed structures in a narrow-gap regime.

Increasing cation intermixing progressively separates these band manifolds. For $\alpha\lesssim0.5$, the localized avoided-crossing minimum is no longer distinguishable because the bands remain separated over the entire $\Gamma$--$Z$ path. Importantly, this evolution does not correspond to a closing and reopening of the global bulk gap. We therefore define $\alpha^{\ast}\approx0.5$ as the characteristic crossover between the narrow-avoided-crossing and enhanced-gap regimes, rather than as a topological critical point. Under these conditions, a symmetry-breaking perturbation, such as the introduction of magnetic atoms (e.g., Mn), the onset of ferroelectric-type structural distortions, or the application of uniaxial strain, may readily drive the system into a Dirac or Weyl semimetal state.

From the perspective of the Fu--Kane parity criterion, the observation that the orbital inversion occurs first at $\Gamma$ alone and later simultaneously at $\Gamma$ and $Z$ with increasing disorder would naively suggest a topological phase transition from $(1;000)$ to $(0;000)$. However, the Wannier charge center (WCC) evolution calculated for the entire series of structures consistently yields the same topological class, namely the strong topological insulator phase $(1;000)$. This apparent contradiction is resolved by recalling that an inversion of the projected orbital weight is an indicator of a parity change of the wave functions only when the relevant states belong to different irreducible representations. If, as is apparently the case at $Z$ for the disordered structures, the band-edge states transform according to the same irreducible representation, the orbital character reversal does not lead to a true parity inversion and therefore does not alter the topological invariant.

\section{Conclusion}

In summary, by combining single-crystal X-ray diffraction, HAADF-STEM, EDX, ARPES, and first-principles calculations, we demonstrate that Ge$_2$Bi$_2$Te$_5$ crystals grown by both chemical vapor transport and the Bridgman method exhibit pronounced cation-sublattice intermixing, closely resembling the behavior previously reported for Ge$_2$Sb$_2$Te$_5$. The degree of cation ordering, characterized by the parameter $\alpha$ defined in Eq.~(1), is found to be slightly lower in vapor-grown crystals, which we attribute to deviations from the ideal Ge/Bi stoichiometry. This observation suggests that cation disorder is not merely an intrinsic property of the material but can also be controlled through the synthesis protocol, in close analogy to Ge$_2$Sb$_2$Te$_5$~\cite{lotnyk2016real-space-8d8}.

Ge$_2$Bi$_2$Te$_5$ occupies a unique position within the GeTe--Bi$_2$Te$_3$ family, close to the boundary separating trivial and topological phases. Owing to the intrinsic $p$-type doping of the material, its topological character cannot be established directly from ARPES measurements alone. However, by combining experimental results with first-principles calculations that explicitly account for cation disorder, we establish the topological nature of Ge$_2$Bi$_2$Te$_5$ through a direct comparison between the measured and calculated electronic band structures \cite{tian2026three-dimensional-228}.

Our central finding is that increasing cation disorder induces a gradual evolution of the electronic structure of Ge$_2$Bi$_2$Te$_5$ from a narrow avoided-crossing gap along the $\Gamma$--$Z$ direction towards an enlarged inverted band gap as the ordering parameter decreases below $\alpha \approx 0.5$, corresponding to about 25\% intermixing of Ge and Bi atoms within the cation sublattice. This pronounced sensitivity of the electronic band structure to cation intermixing establishes disorder as an additional control parameter for tailoring the electronic properties of this material.

Partial density-of-states calculations provide insight into the microscopic origin of this evolution. Increasing cation disorder leads to a redistribution of Ge and Bi atoms within the cation sublattice, modifying the orbital composition of the electronic states near the valence-band maximum. The increased contribution of Bi-derived states is consistent with an enhanced effective influence of spin--orbit coupling, leading to an enlarged inverted band gap, while the global topological indices remain unchanged at $(1;000)$. Since the degree of cation disorder can be tuned during crystal growth, it provides a practical route to engineering the electronic structure within the same topological phase. Thus, using Ge$_2$Bi$_2$Te$_5$ as an example, our findings suggest that controlled cation disorder can serve as an additional degree of freedom for the design and engineering of functional quantum materials.

\section{Acknowledgements} 

A.S.F. acknowledge support of Russian Science Foundation (Grant No. 22-72-10074-P) for the analysis of the ARPES data. A.S.F., L.V.Y. thank the Ministry of Science and Higher Education of the Russian Federation (Grant No. 075-15-2024-632) for the support of the calculations of the bandstructure by means of DFT. 

The authors acknowledge the BESSY II synchrotron for the experimental time granted for ARPES measurements at $1^2$ instrument under proposal number 202-09872-CR.

\section{Experimental Section}
\textbf{Sample Synthesis:} All single crystals were grown from a mixture of 6N pure elemental Ge, Bi and Te. Crystals were prepared from the melt using the Bridgman method and from vapor using chemical vapor transport (CVT). 

Melt growth was performed using a two-phase source (liquid + floating solid tablet) in a Bridgman-like growth configuration as described in Ref.~\cite{sergeev2025growth-648}. In this approach, crystal growth proceeds from a \ch{Bi2Te3} melt equilibrated with solid GeTe. The precursor compounds GeTe and \ch{Bi2Te3} were synthesized by direct alloying of high-purity elemental components in evacuated and sealed quartz ampoules at temperatures exceeding their melting points (780~°C and 620~°C, respectively), followed by melt homogenization for 24~h. The obtained GeTe ingot was ground in an agate mortar and pressed into pellets approximately 8~mm in diameter. A \ch{Bi2Te3} ingot together with GeTe pellets was loaded into a quartz growth ampoule (inner diameter 10~mm) with a conical tip for geometrical selection of spontaneously formed crystals. The ampoule was evacuated, sealed, and placed in a vertical four-zone furnace with a temperature gradient of 6–10~°C/cm at the crystallization front. The temperature at the lower end of the ampoule was maintained at 590~°C, corresponding to the crystallization temperature of \ch{Ge2Bi2Te5} preliminarily determined by differential thermal analysis (DTA). Prior to growth, the ampoule was held under these conditions for 24~h to establish thermal and compositional equilibrium. Crystal growth was carried out by pulling the ampoule at a rate of 0.2~mm/h for 7 days, after which the crystal was quenched in water.

CVT was performed using iodine as a transport agent. The growth charges consisted of 1~g stoichiometric mixtures of elemental Ge, Bi, and Te loaded into a quartz ampoule (10~mm inner diameter, 12–14~cm in length) together with a sealed capillary containing iodine. The ampoule was subsequently evacuated and sealed. The iodine concentration was calculated to give a pressure of approximately 0.3~atm at 580~°C for \ch{I2}. The ampoules were heated in a temperature gradient with the source temperature of 581~°C and the growth temperature of 532~°C for 7 days. The largest obtained single crystals were mechanically separated and selected for further studies.

The obtained crystals were characterized by X-ray fluorescence (XRF) and further analyzed using powder and single-crystal X-ray diffraction (XRD). The XRF measurements were performed using a Bruker Mistral-M1 microfocus system equipped with an XFlash 30~mm$^2$ detector. Concentrations were determined using XSpect software by analyzing the area of individual peaks with the external standard model, with 4–7 measurements taken for each crystal. 

\textbf{Structural Analysis:}

For powder X-ray diffraction (XRD) measurements, the samples were ground into powder. XRD was conducted using a Rigaku SmartLab SE diffractometer equipped with a PSD Pixel3D detector in the $2\theta$ range of $10^\circ$--$80^\circ$ with a step size of 0.02° at room temperature. The obtained data were processed using the Jana2006 software package. 

Single-crystal X-ray diffraction data were collected using a Rigaku XtaLAB Synergy-S diffractometer equipped with a PhotonJet-S detector operating with Mo~K$\alpha$ radiation at 50~kV and 1~mA. A crystal fragment with homogeneous chemical composition was mechanically exfoliated to a thickness below 100~$\mu$m, and a flat region free of macroscopic curvature was chosen for the experiment. For each crystal, more than a hemisphere of data was collected with a frame width of 0.5° in $\omega$, and 9–10~s were spent counting per frame. The data were integrated and corrected for absorption, applying a multi-scan type (ABSCOR3) model using the CrysAlis Pro (Rigaku Oxford Diffraction) software. The structures were refined using the Jana2006 software \cite{petek2014crystallographic-291}. The parameters of the X-ray diffraction experiment and structure refinement are given in Table~S1-S3 (see the supplementary materials). 

For the HAADF-STEM measurements, a cross-section of a single crystal was fabricated using a Helios G4 PFIB uXe dual-beam system with a focused ion beam (FIB) of Xe$^+$ plasma. A protective Pt layer of about 1~$\mu$m total thickness was deposited on the region of interest. The blank was thinned at low accelerating voltages to a thickness of about 100~nm. The resulting lamella was fixed in a double-angle holder and placed in a TEM column. Electron diffraction (ED) and HAADF-STEM images were obtained using a Titan Themis Z transmission electron microscope equipped with a DCOR+ condenser spherical aberration corrector at an accelerating voltage of 200~kV. 

\textbf{ARPES measurements:} 

ARPES data were measured at the UE112-PGM2a undulator using the ARPES one-cubed endstation of the BESSY II synchrotron in Berlin. The endstation was equipped with a hemispherical Scienta R4000 analyzer. The energy and angular resolutions of the ARPES measurements were \SI{45}{\meV} and \SI{0.75}{\degree}, respectively. Prior to measurements, the samples were cleaved \textit{in situ} under ultrahigh vacuum (UHV) at pressures below $1 \times 10^{-8}$~mbar.

\textbf{Calculations:} 

Structural optimization calculations were carried out within density functional theory using the projector augmented-wave (PAW) method \cite{PAW1,PAW2} as implemented in the open-source Quantum ESPRESSO code \cite{QE-2009,QE-2017}. The exchange-correlation energy was described by the PBE version of the generalized gradient approximation \cite{perdew}. The surface band structure was treated using the supercell approach and iterative surface Green's functions (SGF) as implemented in the WannierTools package \cite{wu2018wanniertools-7bf}. The SGFs were constructed using maximally localized Wannier functions (WFs) obtained using the Wannier90 code \cite{pizzi2020wannier90-657}. Structural optimization was performed in fully relativistic mode with spin-orbit interaction. The accuracy of the real-space numerical integration was determined by a cutoff energy of 100~Ry, the total-energy convergence criterion was $10^{-7}$~Ry, whereas the surface Brillouin zone of the supercell was sampled with a $5 \times 5$ mesh of $k$-points. The convergence thresholds for total energy and forces during ionic minimization were $10^{-4}$~Ry and $10^{-3}$~Ry/bohr, respectively.

\bibliography{references}
\end{document}


\title{Supplemental information for the manuscript: Cation-intermixing driven evolution from a narrow avoided crossing to \\ an enlarged inverted gap in Ge$_2$Bi$_2$Te$_5$}

\author{Alexander~S.~Frolov}
\email{frolov.as@mipt.ru}

\affiliation{Moscow Center for Advanced Studies, Kulakova str. 20, Moscow 140829, Russia}
\affiliation{Donostia International Physics Center (DIPC), 20018 Donostia/San Sebasti\'{a}n, Basque Country, Spain}
\affiliation{Lomonosov Moscow State University, Leninskie Gory 1/3, 119991 Moscow, Russia }

\author{Alexander~V.~Fedorov}
\affiliation{Helmholtz-Zentrum Berlin f\"ur Materialien und Energie, Elektronenspeicherring BESSY II, Albert-Einstein-Str. 15, 12489 Berlin, Germany}

\author{Nadezhda~V.~Vladimirova}
\affiliation{Moscow Center for Advanced Studies, Kulakova str. 20, Moscow 140829, Russia}
\affiliation{Lomonosov Moscow State University, Leninskie Gory 1/3, 119991 Moscow, Russia }

\author{Sergey~M.~Aksenov}
\affiliation{Laboratory of Arctic Mineralogy and Materials Sciences, FRC Kola Science Center, 14 Fersman Str. Apatity, Murmansk Region, 184209 Russian Federation}

\author{Mark~A.~Naumov}
\affiliation{Moscow Center for Advanced Studies, Kulakova str. 20, Moscow 140829, Russia}

\author{Savely~V.~Domenko}
\affiliation{Moscow Center for Advanced Studies, Kulakova str. 20, Moscow 140829, Russia}

\author{Artem~A.~Minaev}
\affiliation{Moscow Center for Advanced Studies, Kulakova str. 20, Moscow 140829, Russia}

\author{Vasily~S.~Stolyarov}
\affiliation{Moscow Center for Advanced Studies, Kulakova str. 20, Moscow 140829, Russia}

\author{Denis~V.~Vyalikh}
\email{denis.vyalikh@dipc.org}

\affiliation{Donostia International Physics Center (DIPC), 20018 Donostia/San Sebasti\'{a}n, Basque Country, Spain}
\affiliation{University of the Basque Country UPV/EHU, 20018 Donostia/San Sebasti\'{a}n, Spain}
\affiliation{IKERBASQUE, Basque Foundation for Science, 48013, Bilbao, Spain}

\author{Lada~V.~Yashina}
\affiliation{Moscow Center for Advanced Studies, Kulakova str. 20, Moscow 140829, Russia}
\affiliation{Donostia International Physics Center (DIPC), 20018 Donostia/San Sebasti\'{a}n, Basque Country, Spain}
\affiliation{Lomonosov Moscow State University, Leninskie Gory 1/3, 119991 Moscow, Russia }

\date{\today}
\maketitle
\tableofcontents
\newpage

\section{Details on structural characterization}

\begin{table}[h!]
\centering
\linespread{0.9}\selectfont
\caption{The solution of the structures of crystals Ge$_2$Bi$_2$Te$_5$ grown by directional crystallization (Modified Bridgman method) and vapor crystal growth (chemical vapor transport method, CVT). Crystal data, data collection, and refinement details.}
\label{tab:crystal_data_Ge2Bi2Te5}
\renewcommand{\arraystretch}{0.95}
\setlength{\tabcolsep}{4pt}
\setlength{\aboverulesep}{1pt}
\setlength{\belowrulesep}{1pt}
\begin{tabular}{L{6cm}C{5cm}C{5cm}}
\toprule
\multicolumn{1}{l}{\textbf{Crystal data}} & Modified Bridgman method & CVT \\
\midrule
Chemical formula &
Ge$_{1.99(6)}$Bi$_{2.01(6)}$Te$_5$ &
Ge$_{2.04(7)}$Bi$_{1.96(7)}$Te$_5$ \\
$M_r$ & 1203.2 & 1195.7 \\
Crystal system, space group &
\multicolumn{2}{c}{Trigonal, $P\bar{3}m1$ (No.~164)} \\
Temperature (K) &
\multicolumn{2}{c}{296} \\
$a, c$ (\AA) &
4.2987(5), 17.3444(10) &
4.2893(3), 17.3532(8) \\
$V$ (\AA$^3$) & 277.56(5) & 276.49(3) \\
$Z$ & \multicolumn{2}{c}{1} \\
Radiation type & \multicolumn{2}{c}{Mo $K\alpha$} \\
$\mu$ (mm$^{-1}$) & 49.97 & 49.44 \\
Crystal size (mm) & 1$\times$0.5$\times$0.04 & 0.9$\times$0.6$\times$0.05 \\
\midrule
\multicolumn{3}{l}{\textbf{Data collection}}\\
\midrule
Diffractometer &
\multicolumn{2}{c}{XtaLAB Synergy, Dualflex, HyPix} \\
Absorption correction &
\multicolumn{2}{p{10cm}}{%
\centering\arraybackslash
Multi-scan. $CrysAlis PRO$ 1.171.41.123a (Rigaku Oxford Diffraction, 2022). Empirical absorption correction using spherical harmonics, implemented in SCALE3 ABSPACK scaling algorithm.} \\
\makecell[l]{No. of measured, independent\\and observed $\left[I > 3\sigma(I)\right]$ reflections} &
4789, 493, 439 &
4413, 494, 449 \\
Index range &
\makecell[c]{$-6\leq h\leq6$, $-6\leq k\leq6$,\\$-27\leq l\leq27$}  &
\makecell[c]{$-6\leq h\leq6$, $-6\leq k\leq5$,\\$-25\leq l\leq27$}
 \\
$R_{\mathrm{int}}$ & 0.269 & 0.245 \\
$(\sin\theta/\lambda)_{\max}$ (\AA$^{-1}$) & 0.812 & 0.806 \\
\midrule
\multicolumn{3}{l}{\textbf{Refinement}}\\
\midrule
$R[F^2>2\sigma(F^2)]$, $wR(F^2)$, $S$ &
0.090, 0.105, 6.44 &
0.090, 0.110, 6.57 \\
No. of reflections & 493 & 494 \\
No. of parameters & 17 & 17 \\
\bottomrule
\end{tabular}
\end{table}

\begin{table}[h!]
\centering
\linespread{0.9}\selectfont
\caption{Atomic coordinates, displacement parameters (\AA$^2$) and site-occupancy factors (SOFs) in the structure of Ge$_2$Bi$_2$Te$_5$ crystals grown by Modified Bridgman method and CVT at 296~K obtained from SC XRD data.}
\label{tab:table_coor_disp}
\renewcommand{\arraystretch}{1.05}
\setlength{\tabcolsep}{4.5pt}
\setlength{\aboverulesep}{1.5pt}
\setlength{\belowrulesep}{1.5pt}
\begin{tabular}{c c@{\hspace{4pt}}c c c c c c c}
\toprule
Sample & \multicolumn{2}{c}{Atom} & \makecell[c]{Wyckoff\\site} & SOF & $x$ & $y$ & $z$ & $U_{\rm eq}$ \\
\midrule
\midrule

\multirow{8}{*}{\rotatebox[origin=c]{90}{\makecell[c]{\small Modified Bridgman\\Ge$_{1.99(6)}$Bi$_{2.01(6)}$Te$_5$}}}
& & Te1 & 2$d$ & 1 & 1/3 & 2/3 & 0.42042(13) & 0.0267(6) \\
\cmidrule(lr){2-9}
& & Te2 & 2$d$ & 1 & 2/3 & 1/3 & 0.20427(11) & 0.0192(5) \\
\cmidrule(lr){2-9}
& & Te3 & 1$a$ & 1 & 0 & 0 & 0 & 0.0247(8) \\
\cmidrule(lr){2-9}
& \multirow{2}{*}{\textbf{M1}} & Ge1 & \multirow{2}{*}{2$d$} & 0.699(16) & \multirow{2}{*}{1/3} & \multirow{2}{*}{2/3} & \multirow{2}{*}{0.10390(13)} & \multirow{2}{*}{0.0259(6)} \\
& & Bi1 & & 0.301(16) & & & & \\
\cmidrule(lr){2-9}
& \multirow{2}{*}{\textbf{M2}} & Ge2 & \multirow{2}{*}{2$c$} & 0.294(15) & \multirow{2}{*}{0} & \multirow{2}{*}{0} & \multirow{2}{*}{0.32818(10)} & \multirow{2}{*}{0.0250(4)} \\
& & Bi2 & & 0.706(15) & & & & \\
\midrule
\midrule

\multirow{8}{*}{\rotatebox[origin=c]{90}{\makecell[c]{\small CVT\\Ge$_{2.04(7)}$Bi$_{1.96(7)}$Te$_5$}}}
& & Te1 & 2$d$ & 1 & 1/3 & 2/3 & 0.42046(13) & 0.0268(6) \\
\cmidrule(lr){2-9}
& & Te2 & 2$d$ & 1 & 2/3 & 1/3 & 0.20300(10) & 0.0196(5) \\
\cmidrule(lr){2-9}
& & Te3 & 1$a$ & 1 & 0 & 0 & 0 & 0.0199(7) \\
\cmidrule(lr){2-9}
& \multirow{2}{*}{\textbf{M1}} & Ge1 & \multirow{2}{*}{2$d$} & 0.674(17) & \multirow{2}{*}{1/3} & \multirow{2}{*}{2/3} & \multirow{2}{*}{0.10555(14)} & \multirow{2}{*}{0.0320(7)} \\
& & Bi1 & & 0.326(17) & & & & \\
\cmidrule(lr){2-9}
& \multirow{2}{*}{\textbf{M2}} & Ge2 & \multirow{2}{*}{2$c$} & 0.346(17) & \multirow{2}{*}{0} & \multirow{2}{*}{0} & \multirow{2}{*}{0.32653(10)} & \multirow{2}{*}{0.0255(5)} \\
& & Bi2 & & 0.654(17) & & & & \\

\bottomrule
\end{tabular}

\vspace{1.5em}

\centering
\linespread{0.9}\selectfont
\caption{Anisotropic atomic displacement parameters (\AA$^2$) in the structure of Ge$_2$Bi$_2$Te$_5$ crystals grown by Modified Bridgman method and CVT at 296~K obtained from SC XRD data.}
\label{tab:table_aniso_disp}
\renewcommand{\arraystretch}{1.35}
\setlength{\tabcolsep}{5.5pt}
\setlength{\aboverulesep}{1.5pt}
\setlength{\belowrulesep}{1.5pt}
\begin{tabular}{cccccccc}
\toprule
Sample & Atom & $U_{11}$ & $U_{22}$ & $U_{33}$ & $U_{12}$ & $U_{13}$ & $U_{23}$ \\
\midrule
\midrule

\multirow{5}{*}[-4pt]{\rotatebox[origin=c]{90}{\small \makecell[c]{Modified Bridgman\\Ge$_{1.99(6)}$Bi$_{2.01(6)}$Te$_5$}}}
& Te1 & 0.0336(7) & 0.0336(7) & 0.0128(10) & 0.0168(4) & 0 & 0 \\
\cmidrule(lr){2-8}
& Te2 & 0.0227(6) & 0.0227(6) & 0.0123(9)  & 0.0113(3) & 0 & 0 \\
\cmidrule(lr){2-8}
& Te3 & 0.0246(8) & 0.0246(8) & 0.0248(15) & 0.0123(4) & 0 & 0 \\
\cmidrule(lr){2-8}
& \textbf{M1} & 0.0309(8) & 0.0309(8) & 0.0160(10) & 0.0154(4) & 0 & 0 \\
\cmidrule(lr){2-8}
& \textbf{M2} & 0.0267(6) & 0.0267(6) & 0.0214(7)  & 0.0134(3) & 0 & 0 \\
\midrule
\midrule

\multirow{5}{*}[-4pt]{\rotatebox[origin=c]{90}{\small \makecell[c]{CVT\\Ge$_{2.04(7)}$Bi$_{1.96(7)}$Te$_5$}}}
& Te1 & 0.0294(7) & 0.0294(7) & 0.0217(10) & 0.0147(4) & 0 & 0 \\
\cmidrule(lr){2-8}
& Te2 & 0.0223(6) & 0.0223(6) & 0.0141(8)  & 0.0112(3) & 0 & 0 \\
\cmidrule(lr){2-8}
& Te3 & 0.0214(8) & 0.0214(8) & 0.0170(11) & 0.0107(4) & 0 & 0 \\
\cmidrule(lr){2-8}
& \textbf{M1} & 0.0313(8) & 0.0313(8) & 0.0335(13) & 0.0157(4) & 0 & 0 \\
\cmidrule(lr){2-8}
& \textbf{M2} & 0.0244(6) & 0.0244(6) & 0.0277(8)  & 0.0122(3) & 0 & 0 \\

\bottomrule
\end{tabular}
\end{table}

\clearpage

\section{Transport properties of the crystals}

Bulk single-crystal samples were prepared in a Hall-bar geometry for electrical transport measurements. The length-to-width ratio was $l/d = 1.5$ for both samples, and the thicknesses were $t = 32~\mu\mathrm{m}$ and $56~\mu\mathrm{m}$ for the crystals grown by the modified Bridgman method and chemical vapor transport (CVT), respectively. The relative uncertainty in $l/d$, arising from the finite size of the silver-paste contacts, was estimated to be $5.5\%$. The thickness uncertainty was determined by the micrometer resolution of $1~\mu\mathrm{m}$, corresponding to relative uncertainties of approximately $3\%$ and $1.8\%$ for the Bridgman- and CVT-grown crystals, respectively.


The measurements were performed in a closed-cycle AttoDRY2100 cryostat. The electrical resistance was determined from the slope of the current-voltage (I-V) characteristics measured in a standard four-probe configuration. A Keithley 6221 current source and a Keithley 2182A nanovoltmeter were used for the measurements. The magnetic field was applied parallel to the 
$c$-axis.

\begin{figure}[!htbp]
\includegraphics[width=0.99\linewidth]{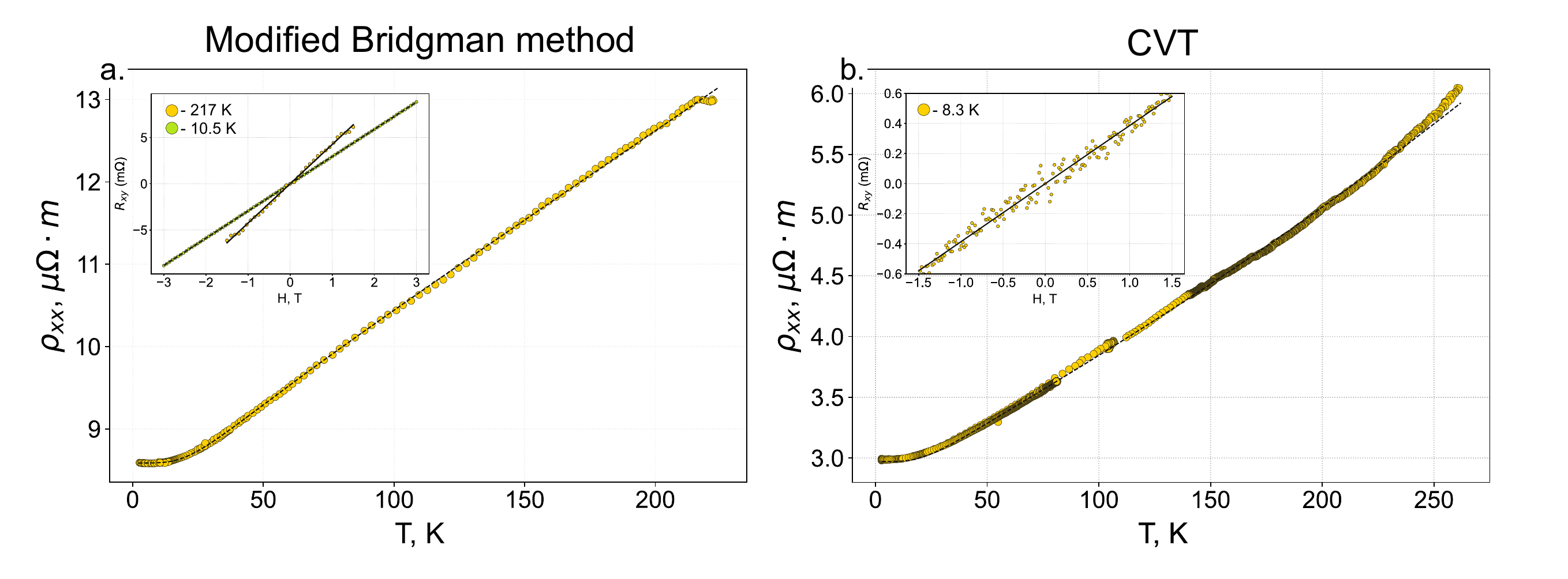}
\caption{\textbf{Electrical transport properties.} Temperature dependence of the longitudinal resistivity $\rho_{xx}$ for single crystals grown by (a) the modified Bridgman method and (b) chemical vapor transport (CVT). Symbols represent experimental data, and dashed black lines show fits to Eq.~\eqref{eq:BG_transport}. Insets: magnetic-field dependence of the Hall resistance $R_{xy}$ measured at (a) $10.5$ and $217~\mathrm{K}$ and (b) $8.3~\mathrm{K}$. Solid black lines show linear fits.}
\label{fig:Transport_225-2}
\end{figure}



The temperature dependence of the longitudinal resistivity was analyzed using the Bloch--Gr\"uneisen expression supplemented by a quadratic term:
\begin{equation}
\label{eq:BG_transport}
\rho_{\mathrm{BG}}(T)
=
\rho_0 + A_{\mathrm{ee}}T^2
+
C\left(\frac{T}{\Theta_D^*}\right)^5
\int_0^{\Theta_D^*/T}
\frac{z^5}{(e^z-1)(1-e^{-z})}\,\mathrm{d}z,
\end{equation}
where $\rho_0$ is the residual resistivity, $A_{\mathrm{ee}}$ is the coefficient of the quadratic contribution, $C$ sets the magnitude of the phonon contribution, and $\Theta_D^*$ is an effective Debye temperature obtained from the transport fit. The agreement with this model is consistent with a phonon contribution to charge-carrier scattering.

For the Bridgman-grown crystal, the Hall carrier concentration varies by approximately $30\%$ over the investigated temperature range, whereas the mobility remains unchanged within the experimental uncertainty.

\section{Topological characterization}


\section{Analysis of the DFT-ARPES correspondence}

\begin{figure*}[h]
\includegraphics[width=0.99\linewidth]{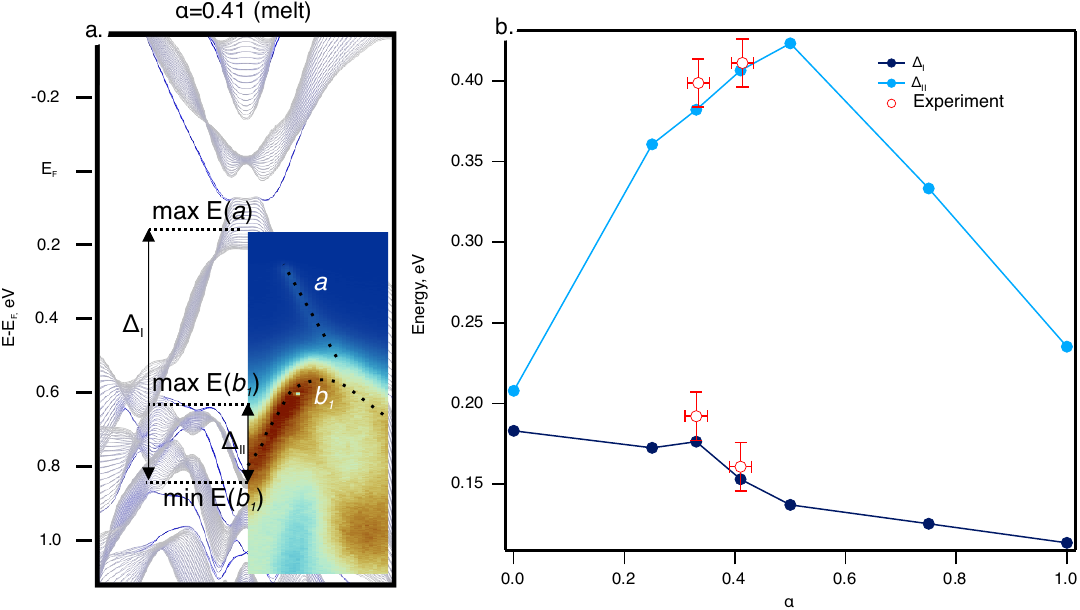}
\caption{\textbf{Quantitative comparison of characteristic energy separations in ARPES and DFT.}
(a) Definition of $\Delta_{\mathrm{I}}$ and $\Delta_{\mathrm{II}}$, illustrated by the calculated band structure at $\alpha=0.41$ and the overlaid ARPES spectrum of the melt-grown crystal. $\Delta_{\mathrm{I}}$ denotes the energy separation between the maximum of the $a$ state and the minimum of the $b_1$ band at $\bar{\Gamma}$, whereas $\Delta_{\mathrm{II}}$ is the difference between the maximum and minimum energies of the $b_1$ band. Dotted lines indicate the relevant dispersions and energy levels. (b) Calculated energy separations as functions of the ordering parameter $\alpha$, compared with the experimental values (open red circles). Horizontal and vertical error bars represent uncertainties in $\alpha$ and the extracted energy separations, respectively.}
\label{fig:ARPES_DFT_comparison}
\end{figure*}

To assess how well the DFT calculations describe the measured electronic structure, we quantitatively compared two characteristic energy separations extracted from ARPES with their calculated dependence on the ordering parameter $\alpha$ [Fig.~\ref{fig:ARPES_DFT_comparison}]. These quantities are defined using the band extrema illustrated in panel (a), with the $b_1$ minimum at $\bar{\Gamma}$ serving as a common reference. The experimental and calculated values show qualitative agreement in both their magnitudes and their variation between the two samples, supporting the calculated evolution of the occupied bands with cation intermixing.

This comparison nevertheless has limitations. In particular, the maximum of the $a$ state is not directly resolved and must be estimated by extrapolating its measured dispersion. Furthermore, the calculated Kohn--Sham band energies are compared with ARPES features that reflect the interacting one-particle spectral function. Finite quasiparticle lifetimes and instrumental resolution broaden these features, introducing uncertainty into the determination of their energy positions. The comparison therefore provides a consistency check of the calculated energy scales.

\bibliography{references}